# Spatial ordering due to hydrodynamic interactions between a pair of colliding drops in a confined shear

Kausik Sarkar[1,2,3] and Rajesh Kumar Singh[2,3]
[1]Department of Mechanical and Aerospace Engineering
George Washington University, Washington, DC 20052
[2]Department of Mechanical Engineering
University of Delaware, Newark, DE 19716, USA
[3]Biomechanics and Movement Science Program
University of Delaware, Newark, DE 19716, USA

PACS numbers: 47.55.D-, 47.57.-s, 47.57.ef, 47.61.Ne

**Abstract**

Pair-collision between viscous drops in a confined shear is numerically simulated to show that the confinement drastically alters the trajectories of the drops. In contrast to free shear, drops here move towards the centerline giving rise to a zero cross-stream separation and a net stream-wise separation. The latter varies as inverse of capillary number and the cube of the confinement (distance between the walls). The stream-wise separation does not depend on the initial positions of the drops. An analytical theory for the phenomenon is offered.

Hydrodynamic interactions between deformable particles and the bounding walls in a confined shear is important in microfluidic applications[1-4] and microcirculatory flows[5]. Due to the small size and velocity, the flow is governed often by the inertia-less Stokes flow. Stokes flow is linear and therefore reversible. A number of counterintuitive phenomena are observed in particulate Stokes flows due to the flow reversibility[6]. For instance, in a free shear, a rigid sphere does not experience any cross-stream motion[7], or a pair of rigid spheres continues in their original streamlines after collision maintaining the pre-collision cross-stream separation. However, for drops, the reversibility is broken; a drop migrates away from a bounding wall[8, 9], and after collision a drop-pair increases their cross-stream separation leading to an enhanced shear induced particle diffusion in an emulsion.[10, 11] Reversibility is also broken in presence of finite inertia [12, 13]. Finite inertia induced particle migration to an intermediate position (0.6 radial distance) in a Poiseuille flow first observed by Segre and Silberberg [14] inspired a series of theoretical and experimental efforts targeted at understanding the underlying physics of inertial migration [15-20].

Recently, we showed that deformation and inertia can work in unison to generate a new type— reversed (type II)—of trajectories for a pair of drops in free shear not seen in Stokes flow [12, 13]. Such reversed trajectories are also seen in presence of inertia for a pair of rigid spheres [21]. The underlying mechanism has been identified as the inertia induced reversed streamlines around a particle [22, 23]. On the other hand, in the Stokes flows limit, in a confined shear a similar reversed (called swapping trajectory by the authors) trajectory for a pair of rigid spheres is discovered due to interaction with the bounding walls [24]. Reversibility leads to a swapping of pre-collision streamlines between drops. Swapping trajectories have been proposed as a probable cause for anomalous particle diffusion observed in an experimental study [25]. Here, we show that in presence of both deformation and confinement, pair interaction gives rise to a specific spatial positioning of drops in the center of the confined domain.

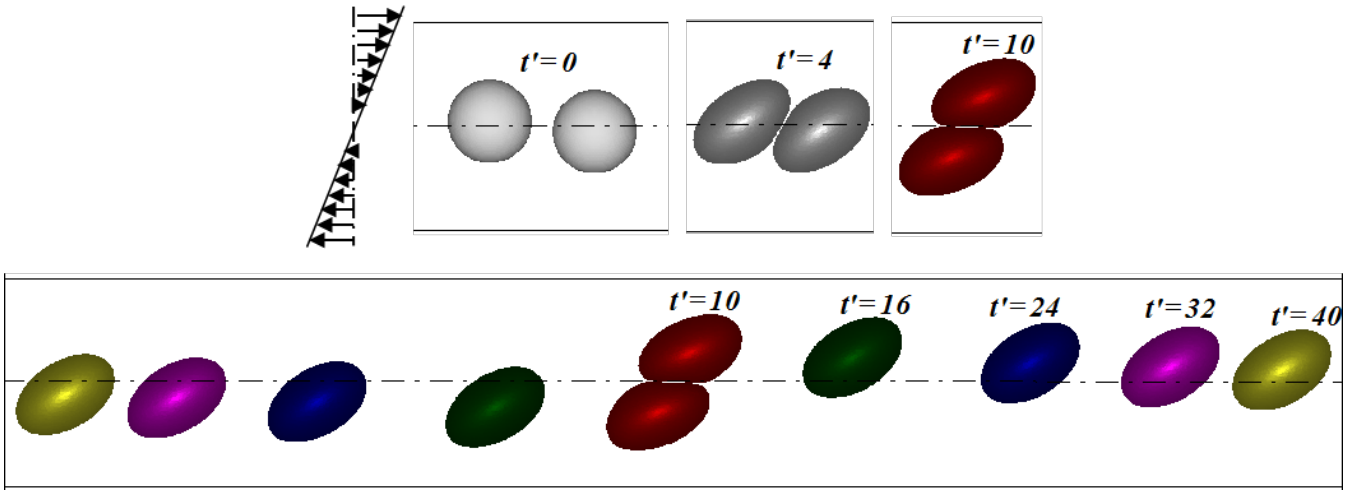

**Figure 1:** Pair of drops interacting in a confined shear for $Ca$=0.2, =5, $\Delta x_o/a$ = 2.5 and $\Delta y_o/a$ = 0.25. Drops travel towards the center of the domain.

We numerically simulate the collision of a pair of initially spherical drops of radius $a$ in a confined shear bounded by walls oriented along the $x$-axis separated by a distance $y = L_y$ using the front tracking finite difference method [26-29]. The method has been used to study a number of different problems, including pair interactions in an unbounded shear, where the simulation showed an excellent match with prior experimental observations[10, 12] The walls are moved with equal and opposite $x$-directional velocity to generate a shear $\dot{\gamma}$. The dynamics depends on the capillary number $Ca = \mu_m \dot{\gamma} a / \Gamma$, viscosity ratio ($\lambda = \mu_d / \mu_m$) and degree of confinement $L_y / a$. Here $\mu_m$ and $\mu_d$ are matrix and drop phase viscosities, and $\Gamma$ is the interfacial tension. Since the code is not fully implicit, we are limited to simulations with small but finite non-zero inertia. We consider Re ($= \rho_m \dot{\gamma} a^2 / \mu_m$) =0.02 as a surrogate for Stokes flow simulation. $\rho_m$ is the density of the matrix phase. We use a computational domain $L_x = 50a$ and $L_z = 5a$. Here, for brevity, we concentrate on a viscosity matched system $\lambda=1$. In the vorticity direction, they are in the same central $z$-plane. We vary $L_y$ to study the effects of confinement on the trajectory of the drops. In the flow ($x$) and the vorticity ($z$) directions periodic boundary conditions are used.

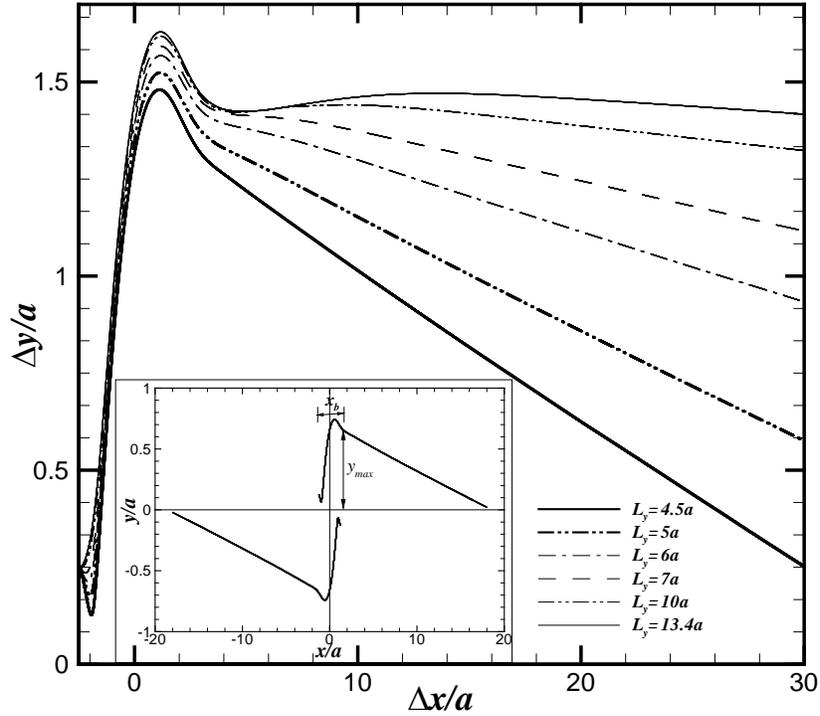

**Figure 2:** Relative trajectory of the drops at $Ca$ = 0.2, $\Delta x_o/a$ = 2.5 and $\Delta y_o/a$ = 0.25 for different $L_y$ values. Inset shows the trajectory of the drops in the domain $L_y$=4.5$a$.

The drops driven by the imposed shear interact, deform—maximum deformation being when they press against each other along the compression axis of the imposed shear—then separate and move in opposite directions (a typical case is shown in Figure 1). However, in contrast to free shear, here after collision drops do not eventually follow any free streamlines [10, 11]. Neither do they achieve a net cross-stream separation. Instead, drops experience a wall induced lateral migration that moves them to the center line, progressively reducing their cross-stream displacement to zero (Figure 2). Finally, they achieve a state of relative equilibrium separated by an equilibrium distance $\Delta x_{final}/a$ at the centerline. This is shown explicitly for the case of $L_y/a = 4.5$ in the inset of Figure 2. For much larger $L_y/a$, i.e. weaker confinement, $\Delta x_{final} / a$ becomes larger, and hence requires much longer simulation in far longer (larger $L_x/a$) computational domain. However, after collision, the drop trajectory eventually becomes a straight line (as can be seen in Figure 2) and therefore, $\Delta x_{final}/a$ can be determined by linear extrapolation. The validity of this extrapolation procedure has been carefully examined and established for several $L_y/a$ by using simulations in longer domains. Only in the limit of very large inter-wall separation ($L_y/a \sim 20$), wall effects are negligible.

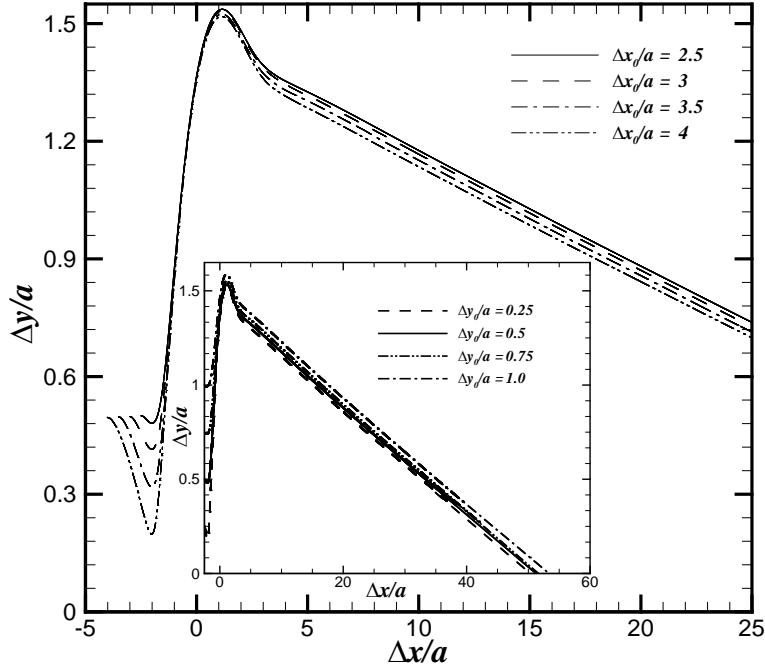

**Figure 3:** Effects of the initial positions on the relative trajectory: Variation of initial separation in the flow direction $\Delta x_0/a$ for $\Delta y_0/a = 0.50$, $Ca=0.20$ and $L_y = 5a$. Inset shows the effect of the separation in the gradient direction $\Delta y_0/a$ for $\Delta x_0/a = 2.50$ in the same domain and the same capillary number.

In Figure 3, we investigate the effects of initial separation on the drop trajectory. Changing initial separation changes trajectory type—increasing initial stream-wise or decreasing initial cross-stream separation leads to reversed or swapping (type II) trajectory both for rigid spheres and drops [12]. However, here we consider those initial positions which do not change the trajectory type. With this restriction, Figure 3 shows that $\Delta x_{final}/a$ remains independent of initial positions. For the results in this paper, we choose initial separation in the flow and the gradient directions fixed at $\Delta x_0 / a = 2.5$ and $\Delta y_0 / a = 0.25$.

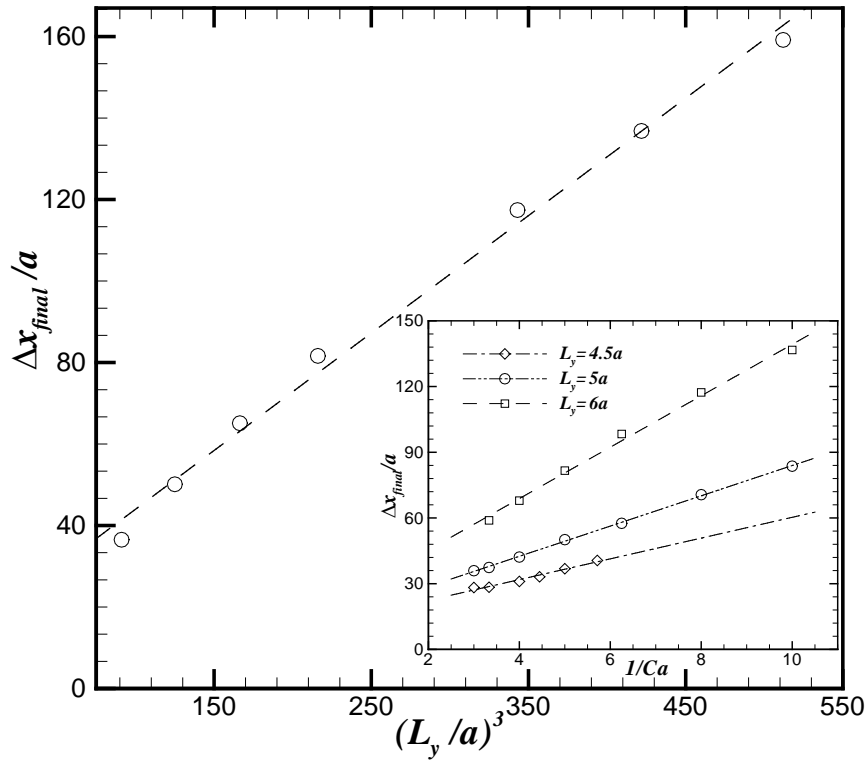

**Figure 4**: Effects of confinement in the gradient direction ($L_y$) on $\Delta x_{final}$ at $Ca=0.20$. Inset of the figure shows that $\Delta x_{final}$ decreases with increasing $Ca$.

Figure 2 studies trajectories for a number of different $L_y / a$, for a capillary number of $Ca = 0.2$ resulting in $\Delta x_{final} / a \sim (L_y / a)^3$ shown in Figure 4. By varying capillary number for three different confinements, we obtain $\Delta x_{final} / a \sim 1/Ca$ shown in the inset of Figure 4. For $Ca > 0.35$ drops experience too large a stretching and possible breakup—confinement is known to delay breakup [30]. They are not considered here.

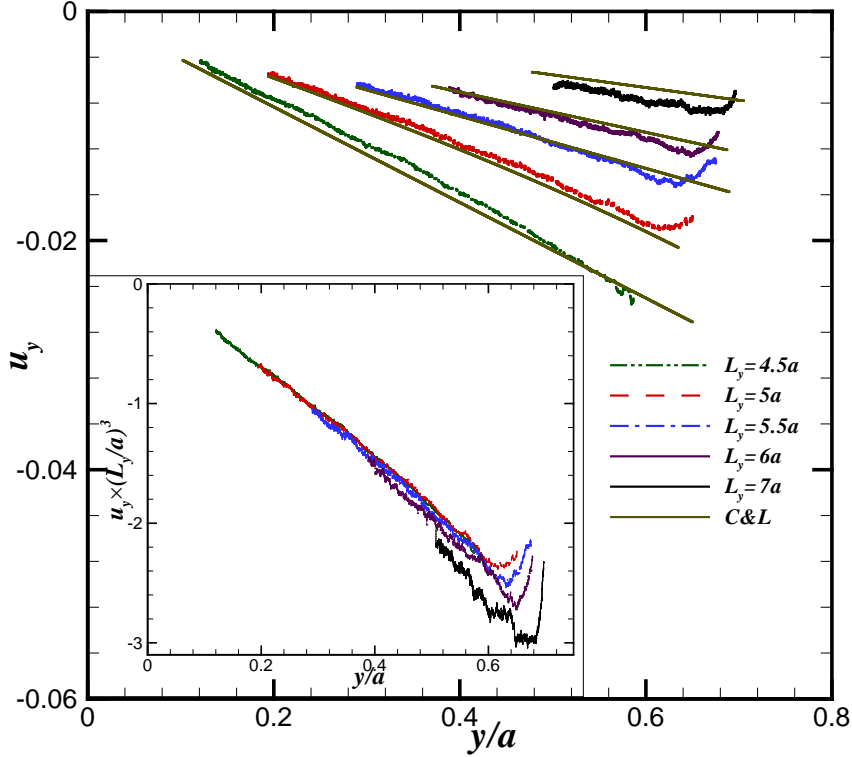

**Figure 5: (Color online)** Variation of lateral velocity of the drops with *y* after collision with increasing confinement from the top along with analytical results due to Chan and Leal (1979). Straight lines Inset shows the scaling of velocity with $L_y$ for different domains.

To investigate the reason for the numerically observed scalings, we investigate the drop velocity. In the flow direction the velocity of the drop post collision is dominated by the imposed shear and therefore can be approximated as $u_x = \dot{\gamma} y$, neglecting the small slip velocity as well as the effect due to the interaction with the other drop, the result becoming more accurate as the drop approaches the centerline. In Figure 5, we note that the lateral velocity $u_y \sim -y$ especially near the centerline after the effects of collision decay. This explains the straight line trajectory of the drop after collision: $dx/dy = u_x/u_y \approx \text{constant}$. Furthermore, inset of Figure 5 shows $u_y \sim -y(a/L_y)^3$. Chan and Leal [31] performed a perturbative analysis of a drop migrating in a plane shear between two parallel plates to get the following migration velocity:

$$\frac{u_y}{\dot{\gamma}a} = \frac{16+19\lambda}{16+16\lambda}\frac{3(54+97\lambda+54\lambda^2)}{70(1+\lambda)^2}Ca\left(\frac{a}{L_y}\right)^2\left(-y*-\frac{8y*}{(1-4y*^2)^2}\right), \tag{1}$$

where $y*$ is the cross-flow distance from the centerline nondimensionalized by $L_y$ ($y* = y/L_y$). For small $y*$ this can be linearized to obtain for $\lambda = 1$:

$$\frac{u_y}{\dot{\gamma}a} = -21.62 \times Ca\left(\frac{a}{L_y}\right)^3\left(\frac{y}{a}\right) \triangleq -\alpha\left(\frac{y}{a}\right) \tag{2}$$

Simulated velocity matches very well this relation in Figure 5 for different $L_y/a$. Noting the symmetry between the top and the bottom drop, one can integrate to obtain

$$\frac{\Delta y}{a} = 2\frac{y_{max}}{a}e^{-\alpha t\dot{\gamma}}, \tag{3}$$

$$\frac{\Delta x}{a} = \frac{\Delta x_b}{a} + \frac{2}{\alpha}\left(\frac{y_{max}}{a}\right)(1-e^{-\alpha t\dot{\gamma}}) = \frac{\Delta x_b}{a} + \frac{2}{\alpha}\left(\frac{y_{max}}{a}\right) - \frac{\Delta y}{a}. \tag{4}$$

Here $\pm y_{max}$ is the post-collision vertical positions of the top and the bottom drops (measured from the centerline) wherefrom they follow a linear trajectory (inset of Figure 2). $\Delta x_b$ is the flow wise separation at that instant. From Figure 2 it seems reasonable and therefore we assume that $y_{max}$ and $\Delta x_b$ are almost independent of $L_y$ and $Ca$. The relation suggested by (4) is verified by the collapse of relative trajectories for different $L_y$ and $Ca$ while scaling $\Delta x$ with $Ca(a/L_y)^3$ in the inset of Figure 6. The same relation (4) after putting $\Delta y_{final}/a = 0$) gives rise to

$$\frac{\Delta x_{final}}{a} = \frac{\Delta x_b}{a} + \frac{2}{21.62 \times Ca}\left(\frac{L_y}{a}\right)^3\left(\frac{y_{max}}{a}\right), \tag{5}$$

explaining $\Delta x_{final}/a \sim (L_y/a)^3/Ca$, if one neglects $\Delta x_b$. Figure 6 shows this scaling for a number of different $L_y/a$ and $Ca$.

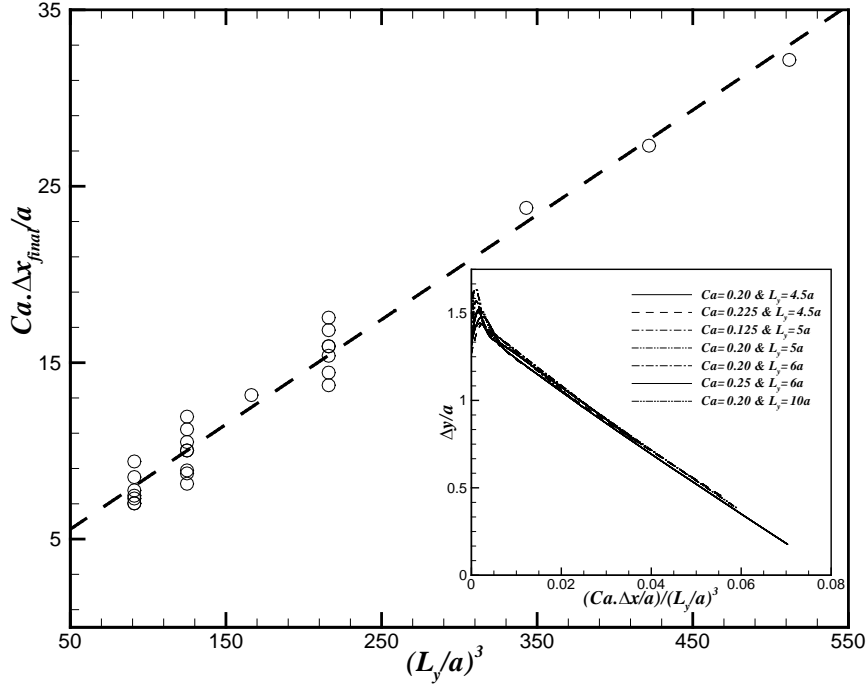

**Figure 6**: Composite scaling of $\Delta x_{final}/a$ with $L_y$ and $Ca$ for different $L_y$ and $Ca$. Inset shows the same for relative trajectory of the drops.

Drops achieving a finite separation ($\Delta x_{final}/a$) in the flow direction in a confined shear is an interesting physical phenomenon analogous to others where particulate system organizes into specific spatial ordering [32]. However, here it is mediated exclusively by hydrodynamic interactions. It assumes further importance in view of the independence of their final separation of the initial drop positions. It indicates that a dilute emulsion of drops in a confined shear has a tendency to organize into a single file separated by a specific distance that would depend on intrinsic hydrodynamic parameters, viz., capillary number and degree of confinement. Note that the parameters studied here are realizable in microfluidic devices. In a 10 micron channel a velocity of 1 cm/s produces a shear rate $\dot{\gamma} \sim 10^3 s^{-1}$; with $\mu \sim$ 1-100 mN/m (water viscosity 1 mN/m), $\Gamma \sim 1\text{-}100$ mN/m/s, for a 2 micron drop ($L_y/a = 5$) capillary number is $Ca \sim 0.00002 - 0.2$ also obtained in microfluidic devices [1, 33]. We have investigated the drop interaction in linear shear instead of in a pressure driven flow, more often used in such devices, because it separates the shear effects on migration from those due to shear gradient present in the latter. The present phenomenon of spatial ordering can be interrogated, e.g. optically, as a means for determining either size or deformability, both parameters affecting capillary number. Differential migration also offers a way of filtering based on the same parameters. There has been a recent surge in innovative applications of size-differentiated inertial migration of rigid particles in pressure driven microfluidic devices for developing sorting, focusing and flow cytometry [34-37]. Deformation provides an additional parameter to control migration and in systems with inertia will create additional migratory effects. Linear chain of droplets separated by a fixed distance have recently seen many novel applications such as determination of the time evolution of reaction kinetics, protein crystallization and concentration indexing using specially designed droplet-pairs [38-40].